\documentclass[aps,prl,twocolumn,groupedaddress,superscriptaddress,showpacs]{revtex4-1}

\usepackage{graphicx,psfrag,color}
\usepackage{amssymb,latexsym,mathrsfs,amsmath,amsthm}
\usepackage[utf8]{inputenc}
\usepackage[normalem]{ulem}

\begin{document}
\def\be{\begin{equation}}
\def\ee{\end{equation}}
\def\bea{\begin{eqnarray}}
\def\eea{\end{eqnarray}}

\title{Ergodicity in randomly perturbed quantum systems}

\author{Stefano Gherardini}
\affiliation{\mbox{LENS, QSTAR, and Department of Physics
and Astronomy, University of Florence,} via G. Sansone 1, I-50019 Sesto Fiorentino, Italy}
\affiliation{INFN, CSDC and \mbox{Department of Information Engineering, University of Florence,} via S. Marta 3, I-50139 Florence, Italy}

\author{Cosimo Lovecchio}
\author{Matthias M. M\"uller}
\author{Pietro Lombardi}
\author{Filippo Caruso}
\author{Francesco Saverio Cataliotti}
\affiliation{\mbox{LENS, QSTAR, and Department of Physics and Astronomy, University of Florence,} via G. Sansone 1, I-50019 Sesto Fiorentino, Italy}

\begin{abstract}
The theoretical cornerstone of statistical mechanics is the ergodic
assumption that all accessible configurations of a physical system
are equally likely. Here we show how such property arises when an
open quantum system is continuously perturbed by an external
environment effectively \textit{observing} the system at random
times while the system dynamics approaches the quantum Zeno regime.
In this context, by large deviation theory we analytically show how
the most probable value of the probability for the system to be in a
given state eventually deviates from the non-stochastic case when
the Zeno condition is not satisfied. We experimentally test our
results with ultra-cold atoms prepared on an atom chip.
\end{abstract}

\date{\today}


\maketitle

\paragraph{Introduction.--}
In a dissipative physical system the total energy is not a preserved
quantity and the system integrability turns out to be broken.
Accordingly, the dynamics trajectories become irregular, covering
densely the whole phase space after a sufficiently long time
interval \cite{Lebowitz_1}. This behaviour connects to ergodicity,
which is studied by analyzing the discrepancy between the time
average of an observable over a very long (but finite) time window
for a single trajectory and its statistical mean averaged over all
system configurations. The mathematical definition of ergodicity was
initially introduced by von Neumann \cite{Neumann,Neumann2}: his
ergodic theorem ensures that only rarely a system observable
deviates considerably from its average value, according to the
statistical mechanics assumption that all attainable configurations
of a system are equally probable. In accordance with von Neumann's
theory, Peres defined ergodicity in quantum mechanics by considering
the equality of the time average of any quantum operator and its
ensemble average \cite{Peres1}. More recently, in Refs. \cite{Asadi,Zhang1}
quantum ergodicity has been examined using the energy
structure of the system, namely its eigenenergies and energy
spacings, while in \cite{Neil} ergodic dynamics has been proved
in a small quantum system consisting of only three superconducting
qubits, as a general framework for investigating non-equilibrium
thermodynamics.

\begin{figure}[t]
\includegraphics[width=0.38\textwidth,angle=0]{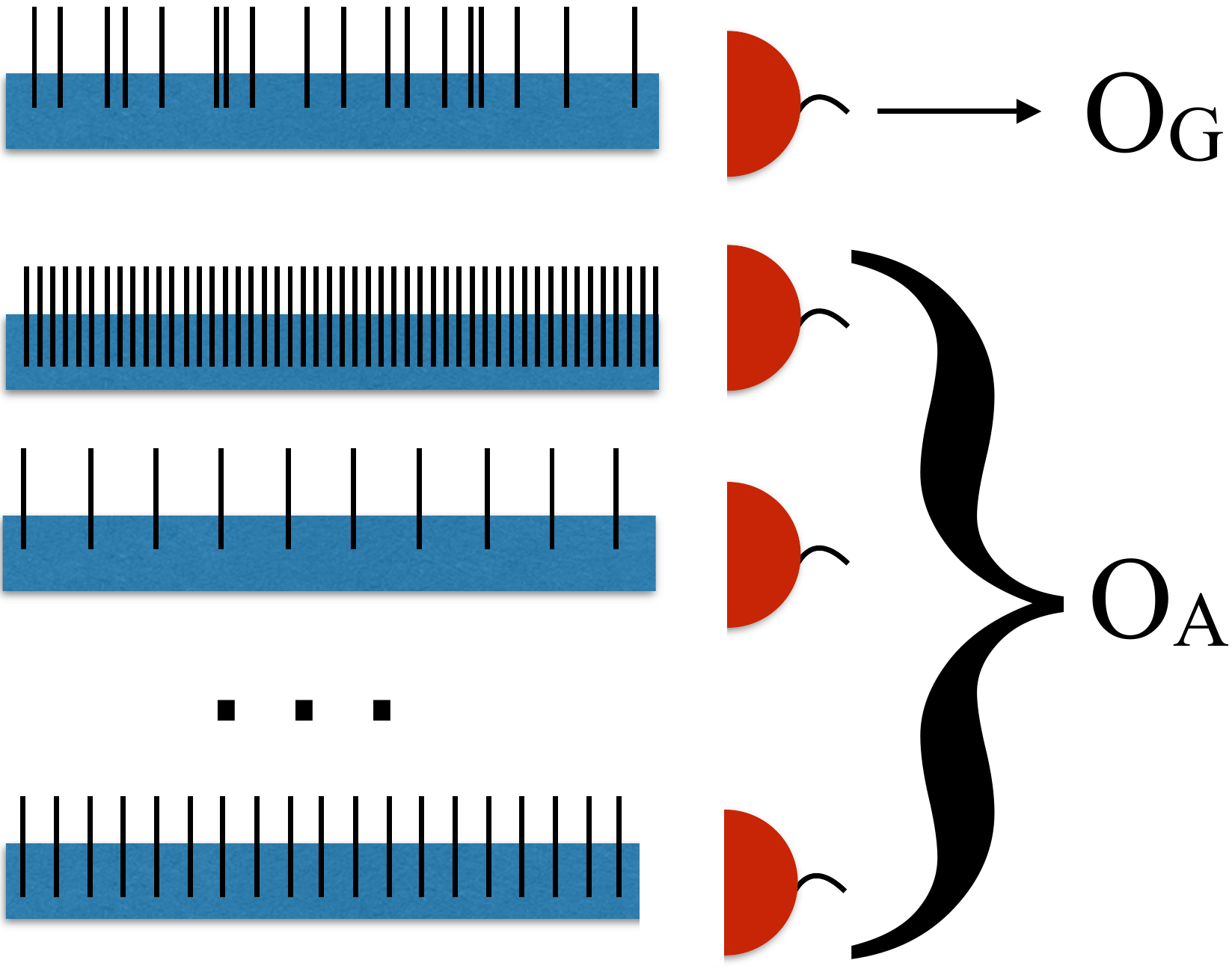}
\caption{Pictorial representation of two measurement schemes on an open quantum system, corresponding, respectively, to here called geometric $O_{G}$ and arithmetic $O_{A}$ averages of a generic observable $O$. The black lines represent the interaction with the external environment that is mimicked by consecutive projective measurements on the system.}
\label{Fig1}
\end{figure}

In a different framework, the theory of large deviations (LD) allows one to
investigate the asymptotic exponential decay of probabilities
associated with large fluctuations of stochastic variables, for both
classical and quantum systems \cite{Ellis1,Touchette1,Dembo1}. The LD
theory has been discussed in the context of quantum gases
\cite{Gallavotti1}, quantum spin systems \cite{Netovcny1} and,
especially, quantum jump trajectories \cite{Garrahan1,Garrahan4} for
open quantum systems \cite{Petruccione1}. In the latter case, the
thermodynamics of a quantum system in continuous interaction with
the environment (external reservoir) has been studied and the
statistics of quantum jumps in the long time limit has been treated
from the point of view of dynamical phase transitions in the
trajectory space. It is worth to emphasize, however, how the
monitoring of general quantum system by a frequent enough series of
measurements is linked to the well-known quantum Zeno effect (QZE)
\cite{Misra1,Pascazio1}, resulting from the statistical indistinguishability 
of neighboring quantum state \cite{Smerzi1}. Very recently, stochastically measured quantum
systems have been introduced and the dependence of Zeno phenomena by
a source of randomness on measurement sequences has been analyzed
\cite{Shushin1}. In particular, in \cite{Gherardini1} a general
quantum system subjected to an arbitrary but fixed number of
consecutive stochastic projective measurements is considered. If the
time interval between the latter is randomly distributed, the
survival probability to remain in the initial quantum state is
itself a random variable taking on different values corresponding to
different realizations of the measurements sequence. This phenomenon
was named stochastic quantum Zeno effect (SQZE).

The main result of this Letter is the theoretical and experimental
demonstration, by applying the LD theory, of the equivalence between
the analogous of time and ensemble averages of the corresponding
stochastic series of quantum projective measurements, hence the
presence of ergodicity, approaching the quantum Zeno regime. This
property shall be useful for the development of new feasible schemes
to control quantum systems by tunable and usually deleterious
stochastic noise.

In this respect, given an observable $O$, we introduce two schemes to take into
account the presence of stochastic noise in terms of an external
environment, applying effectively consecutive quantum projective
measurements to the system, as shown in Fig.
(\ref{Fig1}). We realize these schemes experimentally using a Bose-Einstein
condensate (BEC) of Rubidium ($^{87}$Rb) atoms, trapped on an atom
chip \cite{Schafer}. The first scheme is based on the measurement of
$O$ after the single dynamical evolution of a system that does
interact with the environment at times being stochastically
distributed (geometric average $O_G$). The second scheme does
consist of the averaging of the final observable outcomes over
different dynamical realizations of the system each periodically
interacting with the environment (arithmetic average $O_A$). Then,
different realizations correspond to different time intervals in the
system-environment interaction, but extracted from the same
stochastic probability distribution as in the first scheme. The
reason why they are, respectively, called geometric and arithmetic
averages will be clearer in the following, as well as their relation
with the time and ensemble averages.

\paragraph{Theory.--}

Let us consider a quantum mechanical system under the evolution of a
time-independent Hamiltonian $H$ and initially prepared in a state
$\rho_0=|\psi_0\rangle\langle\psi_0 |$. Then, it is subjected to $m$
consecutive projective measurements $P=|\psi_0\rangle\langle\psi_0|$
after intervals of free evolution of time length $\mu_j$ ($j=1,
\dots, m$).  The observable $O$ is represented by the occupation
number operator of the state $|\psi_0\rangle$, i.e. $P$ itself,
whose measurement here corresponds to the so-called survival
probability, namely, the probability that the system is still in the
initial state at the end of the evolution. We take the $\mu_j$ to be
independent and identically distributed (i.i.d.) random variables
sampled from a given distribution $p(\mu)$, with the normalization
$\int d\mu~p(\mu)=1$ and the mean value $\overline{\mu}$. The
density matrix at the end of the evolution for a total time
$\mathcal{T}\equiv \sum_{j=1}^m\mu_j$, corresponding to a given
realization of the measurement sequence
$\{\mu_j\}\equiv\{\mu_j;~j=1,2,\ldots,m\}$, is given by
$\rho(\{\mu_j\})=\frac{R_{m}(\{\mu_j\})\rho_{0}R_{m}^{\dagger}(\{\mu_j\})}{\mathcal{P}(\{\mu_j\})}$,
where we have defined $R_m(\{\mu_j\})\equiv\prod_{j=1}^{m}PU_{j}P$,
with $U_{j}\equiv\exp\left(-iH\mu_j\right)$, and
$\mathcal{P}(\{\mu_j\})=\rm{Tr}\left[R_{m}(\{\mu_j\})\rho_{0}R_{m}^{\dagger}(\{\mu_j\})\right]$
is the survival probability. The reduced Planck's constant $\hbar$ is set to unity. 
Note that $\mathcal{T}$ is a random
variable that depends on the realization of the sequence
$\{\mu_j\}$, while $\mathcal{P}(\{\mu_j\})$ depends on the system
Hamiltonian $H$ \cite{nota}, the initial density matrix $\rho_{0}$, and also on
the probability distribution $p(\mu)$.

The survival probability can be expressed by $\mathcal{P}(\{\mu_j\})=\prod_{j=1}^{m}q(\mu_{j})$, where $q(\mu_{j})\equiv|\langle\psi_{0}|U_{j}|\psi_{0}\rangle|^{2}$ is the survival probability after a single measurement. By using LD theory, the distribution of $\mathcal{P}$, $\text{Prob}(\mathcal{P})$, can be derived with respect to different realizations of the sequence $\{\mu_{j}\}$ \cite{Touchette1,Gherardini1}. According to the shape of $p(\mu)$, one can calculate the survival probability's average value, i.e.
\be
\langle\mathcal{P}(m)\rangle=\exp\left(m\ln\int_{\mu}d\mu p(\mu)q(\mu)\right) \; ,
\ee
and its most probable value $\mathcal{P}^{\star}$, that is defined as the value at which the rate function $J(\mathcal{P})\equiv\lim_{m\rightarrow\infty}-\ln(\text{Prob}(\mathcal{P}))/m$ takes on its minimum value. The latter is given by
\be\label{4}
\mathcal{P}^{\star}(m) \equiv \exp\left(\int_{\mu}d\mu p(\mu)\ln(q(\mu)^{m})\right)
= \prod_{\{\mu\}}q(\mu)^{mp(\mu)} \; ,
\ee
with the last quantity being the geometric average $\overline{\mathcal{P}}_{g}$ of the survival probability $\mathcal{P}$, weighted with
respect to $p(\mu)$. Additionally, one can consider the ordered case of periodic projective measurements, i.e. $\mu_{j}=\mu$, but with $\mu$ being selected according to $p(\mu)$. This leads to the definition of the arithmetic average, i.e.
\bea\label{6}
\overline{\mathcal{P}}_{a}&=&\int_\mu d\mu p(\mu)q(\mu)^m = \exp\left(\ln\int_{\mu}d\mu p(\mu)q(\mu)^m\right)
\eea
Using Jensen's inequality, namely $\langle\ln(x)\rangle\leq\ln(\langle x\rangle)$, and considering that $\langle x\rangle^{m}\leq\langle x^{m}\rangle$ for any $x \in [0,1]$ and $m\in\mathbb{N}$, it follows that $\overline{\mathcal{P}}_{g}\leq\langle\mathcal{P}\rangle\leq\overline{\mathcal{P}}_{a}$.

As stated, the main result of this Letter is the theoretical demonstration and the experimental observation that it is sufficient to examine the series of constant $\mu$ in order to determine $\overline{\mathcal{P}}_{g}$, $\overline{\mathcal{P}}_{a}$ and $\langle\mathcal{P}\rangle$, even in a \textit{Zeno regime} defined as
\be 
m\int p(\mu)\ln q(\mu)d\mu=m\langle\ln q(\mu)\rangle\ll 1.
\ee
Within this limit, indeed, all these quantities are equal:
\be\label{eq:equal-averages}
\overline{\mathcal{P}}_{a}\approx\left\langle 1+m\ln q\right\rangle = 1 + m\langle\ln q\rangle\approx\overline{\mathcal{P}}_{g}.
\ee
As a consequence of the relation $\overline{\mathcal{P}}_{g}\leq\langle\mathcal{P}\rangle\leq\overline{\mathcal{P}}_{a}$, the equality holds also for the value of the ensemble average. Since in the geometric average (or most probable value) the noise is averaged in time, it turns out that the system dynamics is ergodic. More specifically, we can define the normalized discrepancy $D$ between $\overline{\mathcal{P}}_{g}$ and $\overline{\mathcal{P}}_{a}$ as
\be
D\equiv\frac{\overline{\mathcal{P}}_{a}-\overline{\mathcal{P}}_{g}}{\overline{\mathcal{P}}_{a}} = 1-e^{-\Delta q(\mu,m)}\approx \Delta q(\mu,m),
\ee
where $\Delta q(\mu,m)\equiv\ln\langle q(\mu)^m\rangle - \langle\ln q(\mu)^m\rangle$. Note that $\Delta q(\mu,m)$ is approximately equal to zero in the Zeno regime, otherwise this second-order Zeno approximation breaks down. Indeed, $D$ is not vanishing only at the fourth order in $\mu$, i.e. $\Delta q(\mu,m)\propto\frac{m^{2}}{2}(\Delta^{2}H)^{2}\left(\nu_{4}-\nu_{2}^{2}\right)$, which is determined by the second and the fourth moment of $p(\mu)$ ($\nu_{2}=\int_{\mu}d\mu~p(\mu)\mu^{2}$ and $\nu_{4}=\int_{\mu}d\mu~p(\mu)\mu^{4}$, respectively) and by the variance of the energy, and scales quadratically with the number of measurements $m$ (see SI).
\begin{figure}[t]
\includegraphics[width=0.49\textwidth,angle=0]{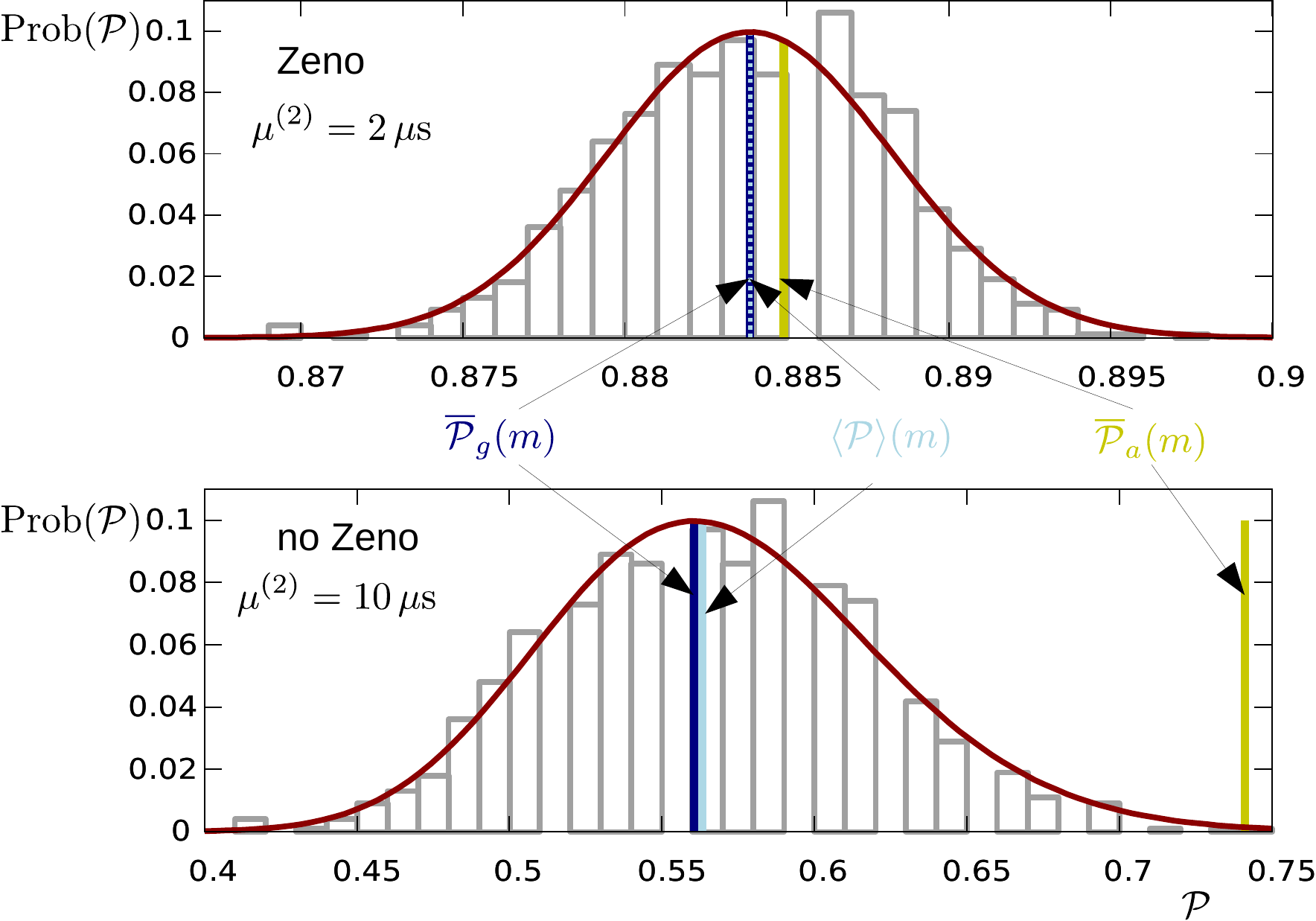}
\caption{Probability distribution Prob($\mathcal{P}$) of the survival probability. The grey boxes refers to the relative frequencies of $\mathcal{P}$, obtained by evaluating numerically $1000$ random realizations of the survival probability after $m=100$ measurements. They are compared to the expected distribution shown in dark red, Eq.~\eqref{eq:prob-distribution1}. The top panel shows the results in the Zeno regime, the lower panel for violated Zeno condition. As it can be seen, the values of the geometric average $\overline{\mathcal{P}}_g$ and of the expectation value $\langle\mathcal{P}\rangle$ are very similar. In the Zeno limit, also $\overline{\mathcal{P}}_a$ is very close to these two values, unlike the lower panel, where the Zeno condition does not hold, validating de facto the ergodic hypothesis for a randomly perturbed quantum system.}
\label{Fig2}
\end{figure}
Fig.~\ref{Fig2} shows the distribution Prob($\mathcal{P}$) of the survival probability.
A bimodal distribution with values $\mu^{(1)}$ and $\mu^{(2)}$ with probability $p_1$ and $p_2$, respectively, is considered for $p(\mu)$.
The grey boxes show the relative frequencies of $\mathcal{P}$ after $m=100$ measurements for $1000$ random realizations as compared to expected distribution shown in dark red, which in the Stirling approximation for $m$ sufficiently large is given by (see SI)
\bea
\mathrm{Prob}(\mathcal{P})\approx\frac{1}{\sqrt{2\pi m p_1 p_2}}\exp\left(-\frac{(k(\mathcal{P})-m p_1)^2}{2m p_1 p_2}\right)\label{eq:prob-distribution1}, \\
k(\mathcal{P})=\frac{\ln \mathcal{P} -m\ln q(\mu^{(2)})}{\ln q(\mu^{(1)})-\ln q(\mu^{(2)})}\label{eq:prob-distribution2}.
\eea

The top panel shows the results for $\mu^{(2)} = 2$  $\mu$s (Zeno condition holds), the lower panel for $\mu^{(2)} = 10$  $\mu$s (Zeno condition violated).
In both scenarios $\overline{\mathcal{P}}_g$ is the maximal value of Prob($\mathcal{P}$) and the expectation value $\langle\mathcal{P}\rangle$ is very close to it. In the Zeno limit, also $\overline{\mathcal{P}}_a$ is very close to these two values, while in the lower panel, where the Zeno condition does not hold, it assumes a distinctly different value, confirming the analytical results. The other parameters are $\Delta H = 2.5$ kHz, $p_{1}=0.8$ and $p_{2}=0.2$. Qualitatively similar behaviours have been observed for other parameter values.

The values of $\overline{\mathcal{P}}_{g}$ and $\overline{\mathcal{P}}_{a}$ can be determined experimentally in the following way. First, we perform $m$ projective measurements on our quantum system with random time intervals $\mu_j$ according to the distribution $p(\mu)$. The value of the survival probability to remain in the initial state that is measured in a single experimental run will very likely be close to $\overline{\mathcal{P}}_{g}$, with deviations that decrease fastly with increasing $m$. On the other hand, the corresponding value of $\overline{\mathcal{P}}_{a}$ can be computed as the average survival probability for $m$ experimental sequences of $m$ projective measurements with the fixed time intervals $\mu_j$ (where in this case the index is referring only to the sequence).

\paragraph{Experimental results.--}
\begin{figure}[h!]
\centering
\includegraphics[width=0.47\textwidth,angle=0]{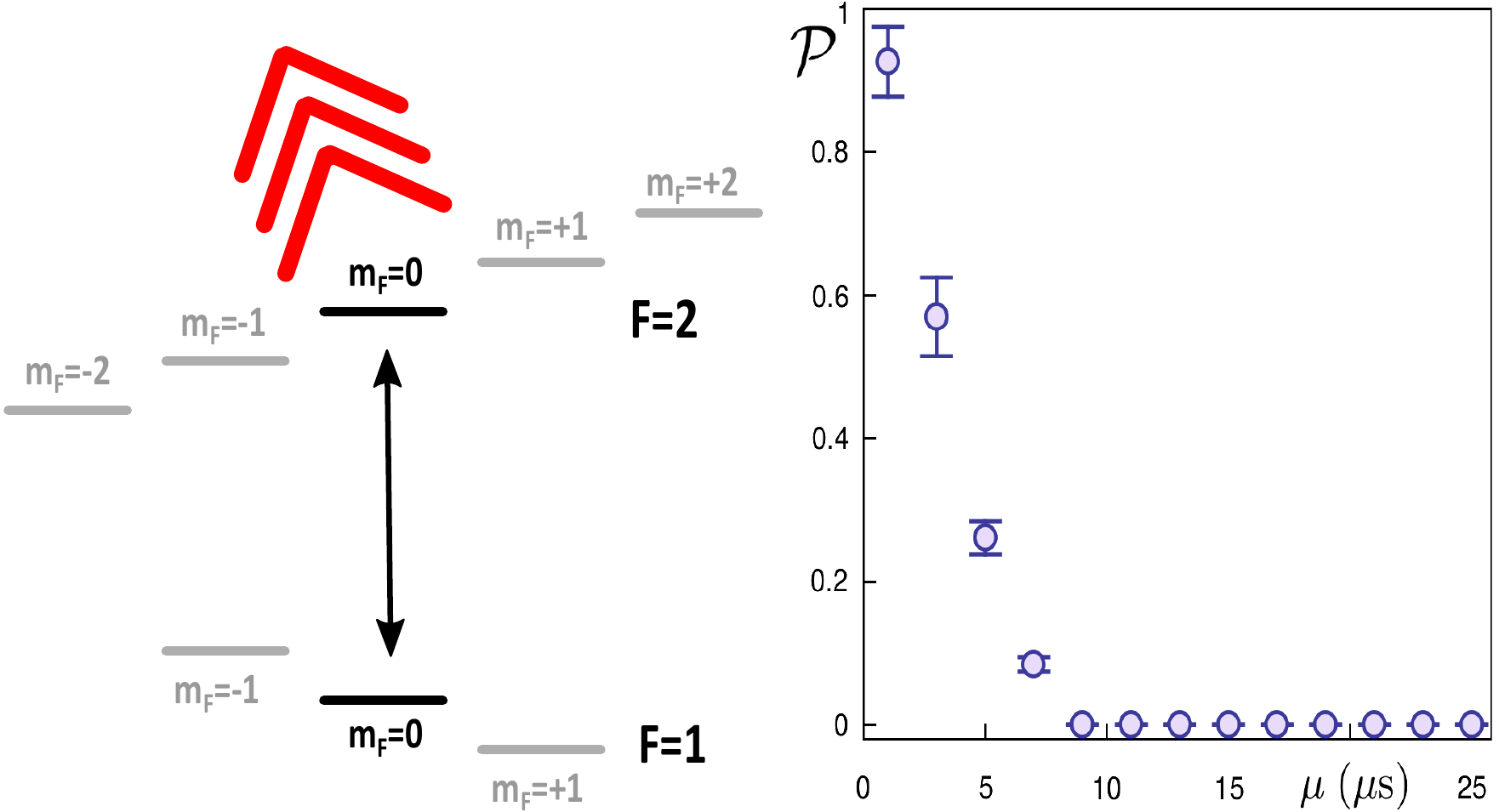}
\caption{Confinement induced by pulsed quantum Zeno effect. The ground state structure of the $^{87}$Rb in presence of a magnetic field consists of two hyperfine levels ($F=1$ and $F=2$), with no internal degeneracy. A laser induced Raman transition couples the sub-levels $|F=1,m_F=0\rangle$ and $|F=2, m_F=0\rangle$, while a laser resonant with the transition $|F=2\rangle\rightarrow|F'=3\rangle$ (red arrows in the picture) deplete the population of the former, and, if strong enough, results equivalent to a projective measurement. On the right we shown the typical exponential decay of the survival probability of the atoms on the $|F=1, m_F=0\rangle$ sublevel while the Raman coupling is on, and simultaneously the resonant light is pulsed 100 times, as a function of the interval $\mu$ between two pulses.} \label{fig_3pr}
\end{figure}
We experimentally test our model with  Bose-Einstein condensed atoms of $^{87}$Rb. In particular the condensate is produced in a magnetic micro-trap realized on an atom chip. We note that the ground state of $^{87}$Rb is a hyperfine doublet separated by $6.834\rm\,GHz$ with total spin $F=2$ and $F=1$ respectively. Our destructive imaging system (see SI) allows, within a single realization, to simultaneously measure the atomic population in each of the 8 possible spin orientations of the two
hyperfine states. After condensation the atoms are released from the magnetic micro-trap and prepared into the the initial state $|\psi_0\rangle=|F=1,
m_F=0\rangle$ in the presence of a homogeneous bias field of $6.179\rm\, G$ (see SI for further information). We allow $0.7 \rm\, ms$ of free fall expansion to suppress the effect of atomic collisions.

The actual experiments are performed by coupling the $|F=1,m_F=0\rangle$ and $|F=2, m_F=0\rangle$ with a Raman transition driven at a Rabi frequency of $5 \rm\, kHz$ by a bi-chromatic laser (Fig.~\ref{fig_3pr}). Since we are working with ground state atoms, with our choice of laser polarizations and thanks to the presence of the homogeneous bias field shifting away from resonance other magnetic sub-levels, we have effectively isolated a closed 2-level system. We realize the projective measurements $P=|\psi_0\rangle\langle\psi_0|$ by shining the atoms with a $1\rm\,\mu s$ pulse of light resonant with the $|F=2\rangle\rightarrow|F'=3\rangle$ component of the Rubidium $D2$ line. Note that from the excited state $|F'=3\rangle$ atoms will immediately decay outside the condensate and will not be seen by our imaging system. Under constant coupling by the Raman beams, we apply $100$ projective measurement $P$ after variable intervals of free evolution $\mu_j$. At the end of the sequence we measure the population remaining in state $|F=1,m_F=0\rangle$ and normalize it by comparison with the population in states $|F=1,m_F=\pm1\rangle$ (see SI for more information). This allows to measure, in a single shot, the survival probability $\mathcal{P}$ of the atoms in the initial state. Each experimental sequence is repeated $7$ times to obtain averages and standard deviations.

\begin{figure}[h!]
\centering
\includegraphics[scale=1.05]{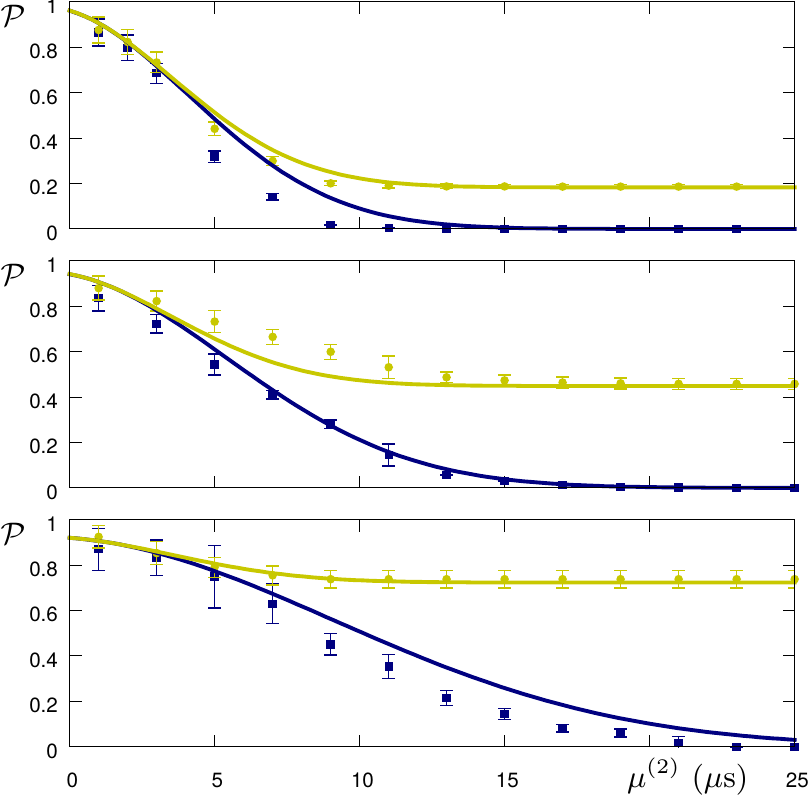}
\caption{Scaling of $\overline{\mathcal{P}}_{g}$ (blue) and
$\overline{\mathcal{P}}_{a}$ (yellow) with the interval length
$\mu^{(2)}$ of a bimodal distribution for $p(\mu)$. In all three
cases $\mu^{(1)} = 2$  $\mu$s, $\Delta H = 2.5$ kHz. The
probabilities of the bimodal distribution $(p_{1},p_{2})$ are $(0.2,
0.8)$ (upper pannel), $(0.5, 0.5)$ (middle pannel), and $(0.8, 0.2)$
(lower pannel). The solid lines are theoretical curves, the single
points are experimental values where the error bars indicate the
standard deviation of the experimental error. The Zeno regime
corresponds to vanishing $\mu_2$.} \label{fig_3}
\end{figure}

In a first set of experiments we fix the intervals of free evolution $\mu_j$ to be all the same and equal to $\mu$ and determine $\mathcal{P}(\mu)$ i.e the probability for an atom to remain in the initial state as a function of $\mu$. As shown in Fig.~\ref{fig_3pr}, $\mathcal{P}(\mu)$ displays the characteristic exponential decay, becoming not evaluable,in our case, after $9\rm\,\mu s$.
Since we are interested in bimodal distributions we take two values of $\mu$, a fixed one $\mu^{(1)}=2\rm\,\mu s$ and $\mu^{(2)}$ variable between $2\rm\,\mu s$ and $25\rm\,\mu s$. After measuring $\mathcal{P}(\mu^{(1)})$ and $\mathcal{P}(\mu^{(2)})$, we then calculate the arithmetic average of the two data with statistical weights $(p_{1},p_{2})$ respectively. In this way we obtain $\overline{\mathcal{P}}_{a}(\mu^{(2)})$ which represents the statistical mean averaged over the two possible system configurations as a function of the variable time $\mu^{(2)}$. In Fig.~\ref{fig_3} we report as yellow dots the results of three choices $(0.2, 0.8)$, $(0.5, 0.5)$, and $(0.8, 0.2)$ for the statistical weights.

In order to determine the time average of a single realization we perform a second set of experiments. In each experimental sequence we now choose the intervals of free evolution $\mu_j$ from a bimodal distribution of $\mu^{(1)}$ and $\mu^{(2)}$ with probabilities $(p_{1},p_{2})$. The results of these experiments correspond to the geometrical average $\overline{\mathcal{P}}_{g}(\mu^{(2)})$ of the the survival probability as a function of $\mu^{(2)}$. We again choose the probabilities $(0.2, 0.8)$, $(0.5, 0.5)$, and $(0.8, 0.2)$  and report the experimental results, as blue squares in Fig.~\ref{fig_3}.
\paragraph{Conclusions and Outlook.--}

We have analytically demonstrated the occurrence of an ergodicity property when a quantum system is perturbed by an external environment that does effectively measure the system at random times with its dynamics approaching a Zeno regime. In particular, we exploit large deviation theory to prove that the most probable value of the probability for the system to be in a given state approaches the non-stochastic case when the Zeno condition holds. Furthermore, we have experimentally tested these results by means of a BEC of Rubidium atoms which are trapped on an atom chip along the transition to the Zeno regime. This will allow one to effectively determine the influence of a stochastic noise source on the dynamics of a controlled quantum system where the unavoidable presence of the environment is instead exploited as a resource. On top of that, although we have explicitly studied the Hamiltonian dynamics when applying the projective measurement on an initial pure state, the method can be also generalized to arbitrary mixed input states with the dynamics being described by completely positive trace-preserving maps \cite{Caruso1} and to the case of measurements projecting the system on higher dimensional Hilbert subspaces, hence leading to Stochastic Quantum Zeno Dynamics \cite{Shushin1,Gherardini1}. Finally, these results are expected to represent further steps towards controlled manipulations of quantum systems via dissipative interactions where one can indeed control the noisy environment or part of it in order to perform desired challenging tasks that are involved in the schemes of future quantum technologies.

\

\begin{acknowledgments}
We acknowledge fruitful discussions with S. Ruffo, S. Gupta, and A. Smerzi.
This work was supported by the Seventh Framework Programme for Research of the European Commission, under the CIG grant QuantumBioTech, and by the Italian Ministry of Education, University and Research (MIUR), under PRIN Grant No. 2010LLKJBX and FIRB Grant Agreements No. RBFR085XVZ and No. RBFR10M3SB.
\end{acknowledgments}

\section{Supplementary Information}

\subsection{Derivation of $\mathrm{Prob}(\mathcal{P})$ for a bimodal $p(\mu)$}

Let us consider a bimodal distribution for $p(\mu)$, with values $\mu^{(1)}$ and $\mu^{(2)}$ and probability $p_{1}$ and $p_{2} = 1 - p_{1}$. The survival probability $\mathcal{P}(\{\mu_{j}\})$, thus, can be written as
\begin{equation*}
\mathcal{P} = q(\mu^{(1)})^{k(\mathcal{P})}q(\mu^{(2)})^{m - k(\mathcal{P})},
\end{equation*}
where $k(\mathcal{P})$ is the frequency of the event $\mu^{(1)}$. By taking the logarithm of $\mathcal{P}$, Eq.~(\ref{eq:prob-distribution2}) follows by resolving for $k(\mathcal{P})$. Moreover, the frequency $k(\mathcal{P})$ is distributed by a binomial probability distribution, namely
\begin{equation*}
\mathrm{Prob}(k(\mathcal{P})) = \frac{m!}{k(\mathcal{P})!(m - k(\mathcal{P}))!}p_{1}^{k(\mathcal{P})}p_{2}^{m - k(\mathcal{P})}.
\end{equation*}
Assuming that for each value of $k(\mathcal{P})$ there exists a single solution $\mathcal{P}$ for Eq.~(\ref{eq:prob-distribution2}), $\mathrm{Prob}(\mathcal{P})$ can be univocally determined from $\mathrm{Prob}(k(\mathcal{P}))$. Hence, by using the Stirling approximation, for $m$ sufficiently large the binomial distribution $\mathrm{Prob}(k(\mathcal{P}))$ can be reasonably approximated to a Gaussian one, obtaining, thus, Eq.~(\ref{eq:prob-distribution1}).

\subsection{Derivation of $\Delta q(\mu,m)$}

Let us consider the series expansion of $q^{m}$ and its logarithm up to the fourth order, namely
\begin{equation*}
q^{m}=1-m\Delta^{2}H\mu^{2}+\frac{m}{12}\left[\gamma_{H}+3(2m-1)(\Delta^{2}H)^{2}\right]\mu^{4}+\mathcal{O}(\mu^{6})
\end{equation*}
and
\begin{equation*}
\ln q^{m}=-m\Delta^{2}H\mu^{2}+\frac{m}{12}\left[\gamma_{H}-3(\Delta^{2}H)^{2}\right]\mu^{4}+\mathcal{O}(\mu^{6}),
\end{equation*}
where $\gamma_{H}\equiv\overline{H^{4}}-4\overline{H^{3}}\overline{H}+6\overline{H^{2}}\overline{H}^{2}-3\overline{H}^{4}$ is the kurtosis of the Hamiltonian. Hence, by neglecting from $\Delta q(\mu,m)$ the higher order terms, it holds that
\bea
\Delta q &\propto& m\Delta^{2}H\nu_{2}-\frac{m}{12}\left[\gamma_{H}-3(\Delta^{2}H)^{2}\right]\nu_{4}\nonumber \\
&+&\ln\left\{1-m\Delta^{2}H\nu_{2}+\frac{m}{12}\left[\gamma_{H}+3(2m-1)(\Delta^{2}H)^{2}\right]\nu_{4}\right\}\nonumber \\
&=& \frac{m^{2}}{2}(\Delta^{2}H)^{2}\nu_{4}-\frac{m^{2}}{2}(\Delta^{2}H)^{2}\nu_{2}^{2}. \nonumber
\eea

\subsection{Imaging and manipulation of the $^{87}$Rb Bose-Einstein condensate.}

We produce a Bose-Einstein condensate (BEC) of $^{87}$Rb atoms in a
magnetic micro-trap realized with an Atom chip. The trap has a
longitudinal frequency of $46~{\rm Hz}$, the radial trapping
frequency is $950~{\rm Hz}$. The atoms are evaporated to quantum
degeneracy by ramping down the frequency of a radio frequency (RF)
field. The BEC has typically $8\cdot10^4$ atoms, a critical
temperature of $0.5~\mu{\rm K}$ and is at $300~\mu{\rm m}$ from the
chip surface. The magnetic fields for the micro-trap are provided by
a Z-shaped wire on the Atom chip and an external pair of Helmholtz
coils. The RF fields for evaporation and manipulation of the Zeeman
states are produced by two further conductors also integrated on the
Atom chip.
\begin{figure}[h!]
\centering
\includegraphics[width=0.40
\textwidth,angle=0]{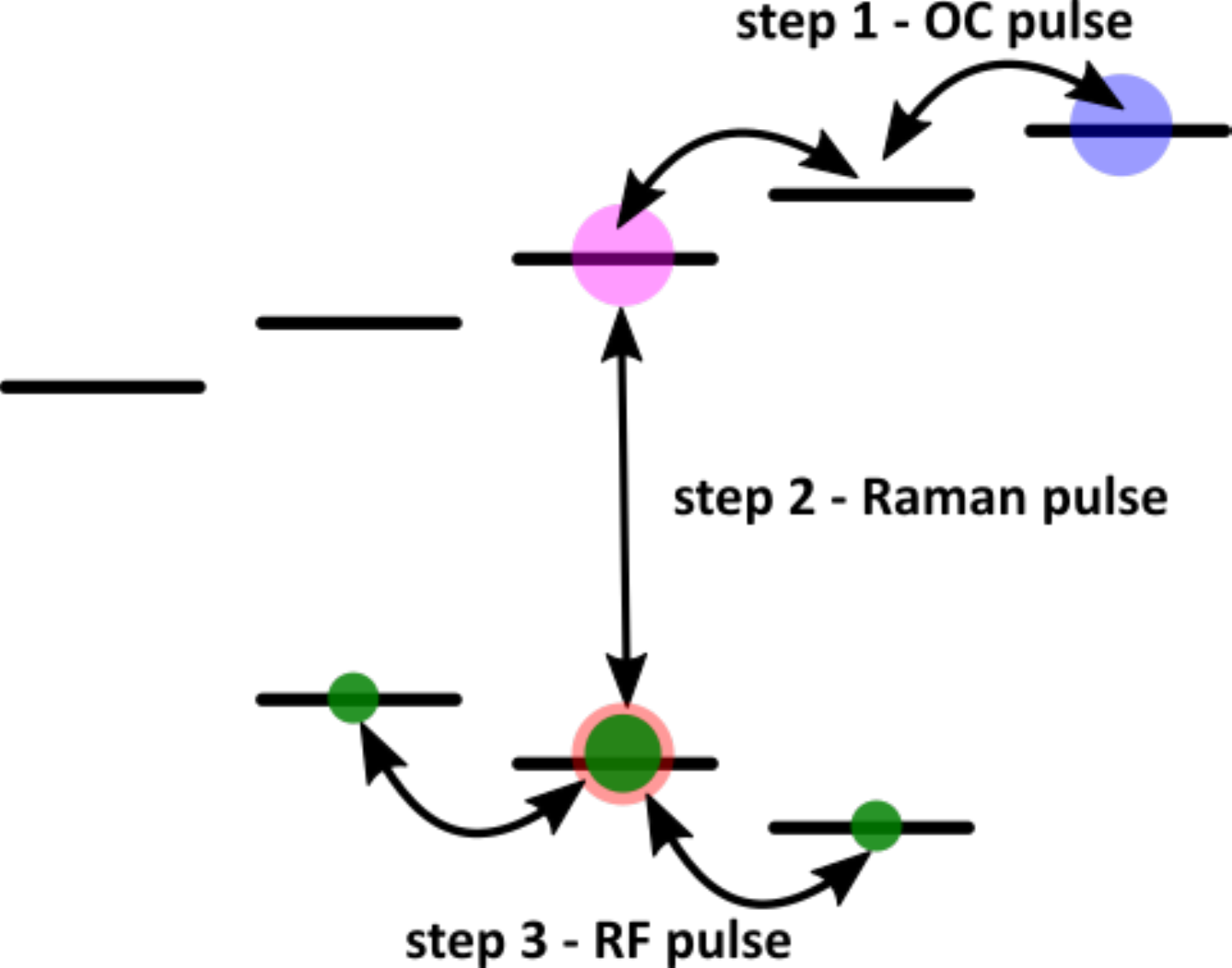}
\caption{State preparation sequence for the SQZE experiment. After the condensation in the pure state $|F=2, m_F=+2\rangle$, in the first step the atoms are transferred, with fidelity higher than $90\%$, in the state $|F=2, m_F=0\rangle$. In the second step the atoms in this sub-level are transferred, by the Raman lasers, in the lower state $|F=1, m_F=0\rangle$, which is the initial state $\rho_0$ for our experiment. In the third, last, step, a fixed amount of population is transferred in the side sublevels $|F=1, m_F=\pm 1\rangle$. These atoms will be used as a benchmark to calculate the survival probability after the experiment.}\label{fig_5}
\end{figure}
To record the number of atoms in each of the 8 $m_F$ states of the
$F=2$ and $F=1$ hyperfine state we apply a Stern-Gerlach method.
After $2~{\rm ms}$ of free fall expansion, an inhomogeneous magnetic field is applied along the
quantization axis for $10~{\rm ms}$. This causes the different $m_F$
sub-levels to spatially separate. After a total time of $23~{\rm ms}$ of
expansion we shine $200 \rm\,\mu s$ of light resonant with the $|F=2\rangle\rightarrow|F'=3\rangle$ pushing away all atoms in the $F=2$ sub-levels and recording the shadow cast by these atoms onto a CCD camera. 
ìWe let the remaining atoms expand for further $1 \rm\,ms$ and then we apply a bi-chromatic pulse containing light resonant to the $|F=2\rangle\rightarrow|F'=3\rangle$ and $|F=1\rangle\rightarrow|F'=2\rangle$ transitions effectively casting onto the CCD the shadow of the atoms in the $F=1$ sub-levels. Another two CCD images to acquire the background complete the imaging procedure.

To prepare the atoms for the experiment, the condensate is released from the magnetic trap and allowed to expand freely for $0.7 \rm\,ms$ while a constant magnetic field bias of $6.179 \rm\,G$ is applied in a fixed direction. This procedure ensures that the atom remain in $|F=2, m_F=+2\rangle$ state and strongly suppresses the effect of atom-atom interactions by reducing the atomic density. The first step of preparation consists in applying a frequency modulated RF pulse designed with an optimal control strategy \cite{Lovecchio} in order to prepare, with high fidelity, all the atoms in the $|F=2, m_F=0\rangle$ state (see Fig.~\ref{fig_5}). This RF optimal pulse is $50\rm\, \mu s$ long. After the RF pulse we transfer the whole $|F=2, m_F=0\rangle$ population in the $|F=1, m_F=0\rangle$ sublevel by shining the atoms with bi-chromatic (Raman) laser light. This is the initial state $\rho_0$ for our experiment. The preparation is completed by applying another RF pulse to place some atomic population in the $|F=2, m_F=\pm 1\rangle$ states for normalization. This atoms will be not altered during the actual experiment, when only the Raman lasers and the resonant light are on, so they can be used as a control sample population. Comparing the fraction of atoms in the $|F=1, m_F=0\rangle$ sub-level survived after the SQZE experiment, with the fraction in the same level without any further manipulation other than the preparation sequence, we are able to retrieve the survival probability $\mathcal{P}$.


\begin{thebibliography}{99}
\bibitem{nota}
Given the initial state $|\psi_{0}\rangle$, we define the expectation value of the $k-th$ power of the system Hamiltonian $H$ as $\langle H^{k}\rangle\equiv\langle\psi_{0}|H^{k}|\psi_{0}\rangle$, $k = 0,1,\ldots$, and 
$\Delta^{2}H\equiv\langle H^{2}\rangle - \langle H\rangle^{2}$ as the variance of the energy.
\end{thebibliography}
\end{document}